\documentclass[usenatbib]{mn2e}
\usepackage{natbib}
\usepackage{times}
\usepackage{graphicx}

\def\oversim#1#2{\lower0.5pt\vbox{\baselineskip0pt \lineskip-0.5pt
     \ialign{$\mathsurround0pt #1\hfil##\hfil$\crcr#2\crcr\sim\crcr}}}

\newcommand{\lpup}{L$_2$~Pup}
\newcommand{\lpuppis}{L$_2$~Puppis}

\title[The light curve of L$_2$~Pup]
{The light curve of the semiregular variable L$\bf_2$~Puppis: I. A recent
dimming event from dust} 
\author[T.R. Bedding et al.]
       {T.~R.~Bedding,$^1$\thanks{E-mail: \tt bedding@physics.usyd.edu.au}
       	A.A.~Zijlstra,$^2$
	A.~Jones$^3$, 
        F.~Marang$^4$,
	M.~Matsuura$^2$,
        A.~Retter$^1$, \newauthor
	P.A.~Whitelock$^4$ and
	I.~Yamamura$^5$ \\
	$^1$School of Physics, University of Sydney 2006, Australia\\
	$^2$UMIST, Department of Physics, P.O. Box 88, Manchester M60 1QD, UK\\
	$^3$Carter Observatory, P.O. Box 2909, Wellington, New Zealand\\
	$^4$SAAO, P.O. Box 9, Observatory 7935, South Africa\\
        $^5$The Institute of Space and Astronautical Science, 
            3-1-1 Yoshino-dai,  Sagamihara, 299-8510, Kanagawa, Japan
}
\pubyear{2002}

\begin{document}

\maketitle

\begin{abstract}

The nearby Mira-like variable \lpup{} is shown to be undergoing an
unprecedented dimming episode.  The stability of the period rules out
intrinsic changes to the star, leaving dust formation along the line of
sight as the most likely explanation.  Episodic dust obscuration events are
fairly common in carbon stars but have not been seen in oxygen-rich stars.
We also present a 10-$\mu$m spectrum, taken with the Japanese IRTS
satellite, showing strong silicate emission which can be fitted with a
detached, thin dust shell, containing silicates and corundum.

\end{abstract}


\section{Introduction}

\lpuppis{} (HR 2748; HIP 34922) is a bright nearby red giant with a
pulsation period of about 140\,d.  Its spectral type of M5eIII and
luminosity of 1500 L$_\odot$ indicate that it is evolving towards the tip
of the Asymptotic Giant Branch (AGB).  Evidence for mass loss at a rate of
$3 \times 10^{-7}\rm \, M_\odot \, yr^{-1}$ \citep{JCP2002} supports this.
\lpup{} is possibly the nearest star in this evolutionary phase, at a
Hipparcos distance of $61 \pm 5$\,pc.  Among known long-period AGB stars,
only R~Doradus has a similar distance.  At 12 microns, \lpup{} is among the
15 brightest sources in the IRAS point source catalogue.

\lpup{} is unusual in several respects.  Firstly, it shows a high degree of
optical polarization, with a variable wavelength dependence that implies a
long timescale for the growth and dissipation of dust grains (of the order
of a decade; \citealt{MCLG86}).  Secondly, CO measurements by \citet{K+O99}
indicate a very low expansion velocity (about 2.5\,km\,s$^{-1}$), which led
them to label \lpup{} as an extreme case, with one of the smallest
expansion velocities ever measured for an AGB star.  The slow wind from
\lpup{} led \citet{WLBJ2000,WLBN2002} to suggest that this star could
represent their B-model, in which mass loss is driven entirely by
pulsations, without any significant input from radiation pressure on dust
grains.  This has been further discussed by \citet{JCP2002}, who modelled
the mass loss and suggested that the pulsations may be non-radial.

Thirdly, as we report here, this star has shown a remarkable change in mean
visual magnitude over the past century, and is currently undergoing a
dramatic dimming.  We present visual and infrared photometry which
characterizes this behaviour, and argue that the most likely cause is the
formation of dust along the line of sight.  We also present the first
10-micron spectrum of \lpup, obtained with the Japanese IRTS satellite,
which shows strong silicate emission.

\section{Light curve}

\subsection{Visual observations}

\begin{figure*}
\includegraphics[width=\textwidth]{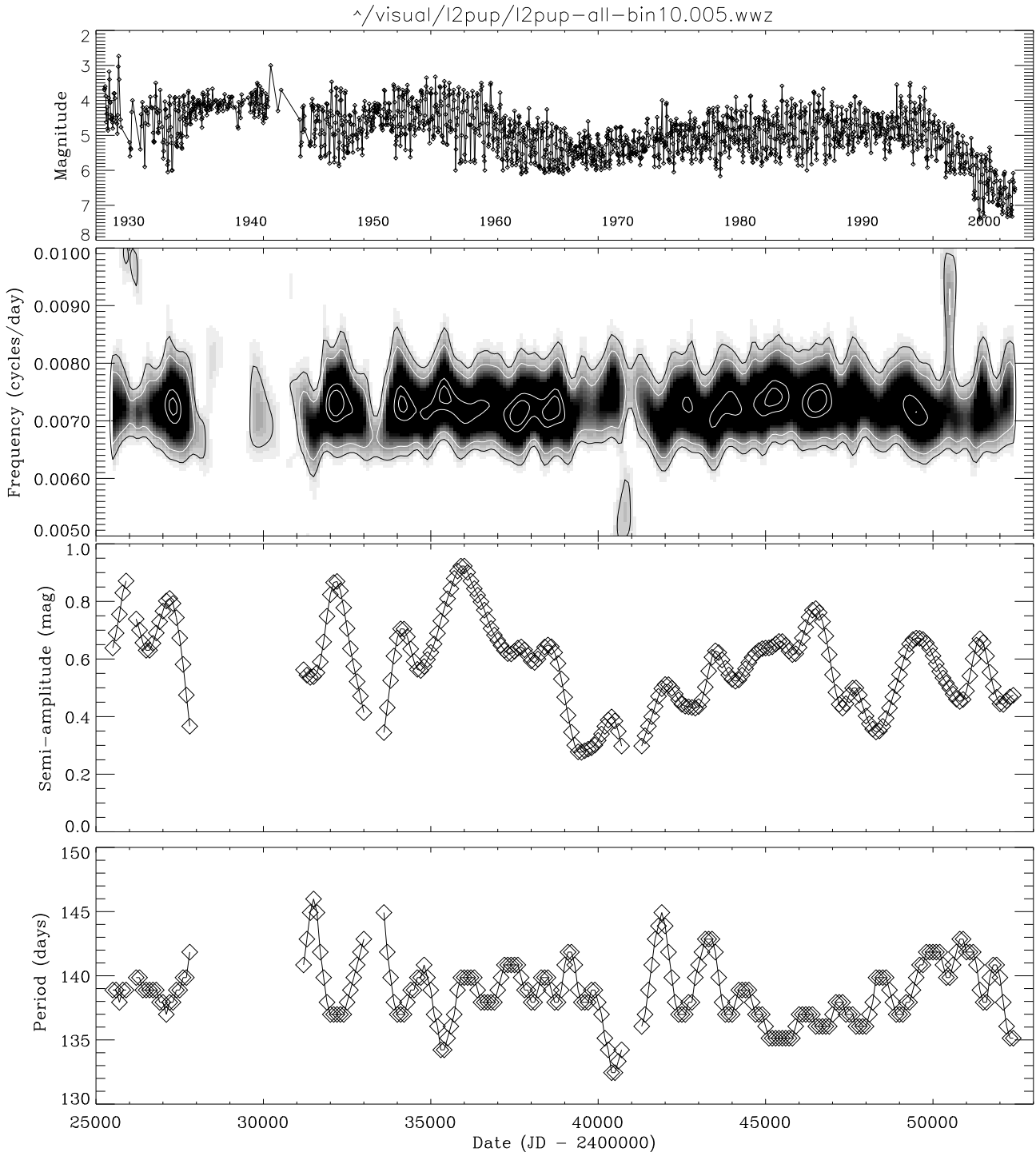}
\caption{\label{fig.wavelet} Wavelet analysis for \lpup.  Top to bottom:
the light curve (in 10-day bins), the WWZ transform, the semi-amplitude and
the period.  }
\end{figure*}

We have analysed visual observations of \lpup{} from the following sources:
the Royal Astronomical Society of New Zealand (RASNZ; 12100 measurements by
110 observers, including 1100 by A.~Jones), the Variable Star Observers
League in Japan (VSOLJ; 792 measurements by 6 observers) and the
Association Francaise des Observateurs d'Etoiles Variables (AFOEV; 389
measurements by 3 observers).  Only data from individual observers
contributing 30 or more observations were used and we did not attempt to
correct for offsets between observers.  The top panel of
Fig.~\ref{fig.wavelet} shows the combined data, binned to 10-day averages.

The variation in mean magnitude is evident.  Before 1945 the data are too
patchy and/or noisy to reach strong conclusions.  However, it is clear that
the star faded significantly after 1960, recovering around 1975 but
remaining fainter by about 0.5 mag compared to the 1950s.  Recently,
starting around 1994, a dramatic fading has occurred.  Even at maximum, the
star now remains fainter than magnitude~6, and it is no longer a naked-eye
variable.

We have used wavelet analysis to search for period and amplitude changes.
Wavelets have been used previously to study long-period variables (e.g.,
\citealt{SGK96,BZJ98,KSC99}).  We used the weighted wavelet Z-transform
(WWZ; \citealt{Fos96}), developed specifically for unevenly sampled data.
We experimented with different values for the parameter $c$, which defines
the tradeoff between time resolution and frequency resolution \cite{Fos96},
and settled on $c=0.005$ as a good compromise.  More details of the
application of the WWZ transform to long-period variables are given by
\citet{BZJ98}.

The lower three panels of Figure~\ref{fig.wavelet} show the wavelet plots
for \lpup, based on the light curve in the top panel.  The second panel
shows the WWZ transform, with the grey scale indicating the significance of
each frequency as a function of time (see \citealt{BZJ98}).  Only a small
range of frequencies is shown -- there is no evidence for significant
power outside this range.  The third and fourth panels show, for each time
bin, the semi-amplitude (in magnitudes) and period (in days) corresponding
to the peak of the WWZ in the second panel.

It is clear from Fig.~\ref{fig.wavelet} that the period of \lpup{} has
remained within a fairly narrow range (135--145\,d) for the past 75 years.
Such period jitter is common among semiregular and Mira variables.  There
is a slight indication of a lengthening of period {\em before\/} the onset
of the recent dimming, from 136 to 142 days, but the change (at most 4\%)
is well within the range of normal Mira variables.  The amplitude, on the
other hand, has changed significantly over the years, but the changes are
apparently uncorrelated with the period jitter.

\lpup{} is classified as an SRb variable, a class of stars with poorly
defined periodicity.  In fact, the period of this star is remarkably
stable.  The characteristics of the wavelet plot show that \lpup{} should
be classified as SRa, a class closely related to Miras, often differing
only in having smaller amplitudes.  

\subsection{Early observations}

\lpup{} was discovered to be variable by Gould in 1872 \citep{C+P07}.
\citet{Rob1893} estimated a period of 137.2\,d from the dates of nine
maxima observed by himself, Gould and Williams during 1872--92.
\citet{C+P07}, in the {\em Second Catalogue of Variable Stars}, quoted a
period of 140.15\,d, which they attributed to Roberts.  \citet{C+P09}
listed dates of maxima going back to 1872, and also adopted a period of
140.15\,d.  It therefore seems that the period of \lpup{} has remained
within a narrow range since its discovery.

The visual magnitude at maximum during the late 1800s was about 3.6
\citep{Wil1897,C+P09}, consistent with modern (1950s) out-of-decline
values. \citet{M+H22} listed the magnitude range since discovery as
3.4--4.6 at maximum and 5.8--6.2 at minimum.  This suggests that, between
1872 and 1918, \lpup{} was never as faint as it currently is.

\subsection{Near-infrared photometry}

\begin{figure*}
\includegraphics[width=\textwidth]{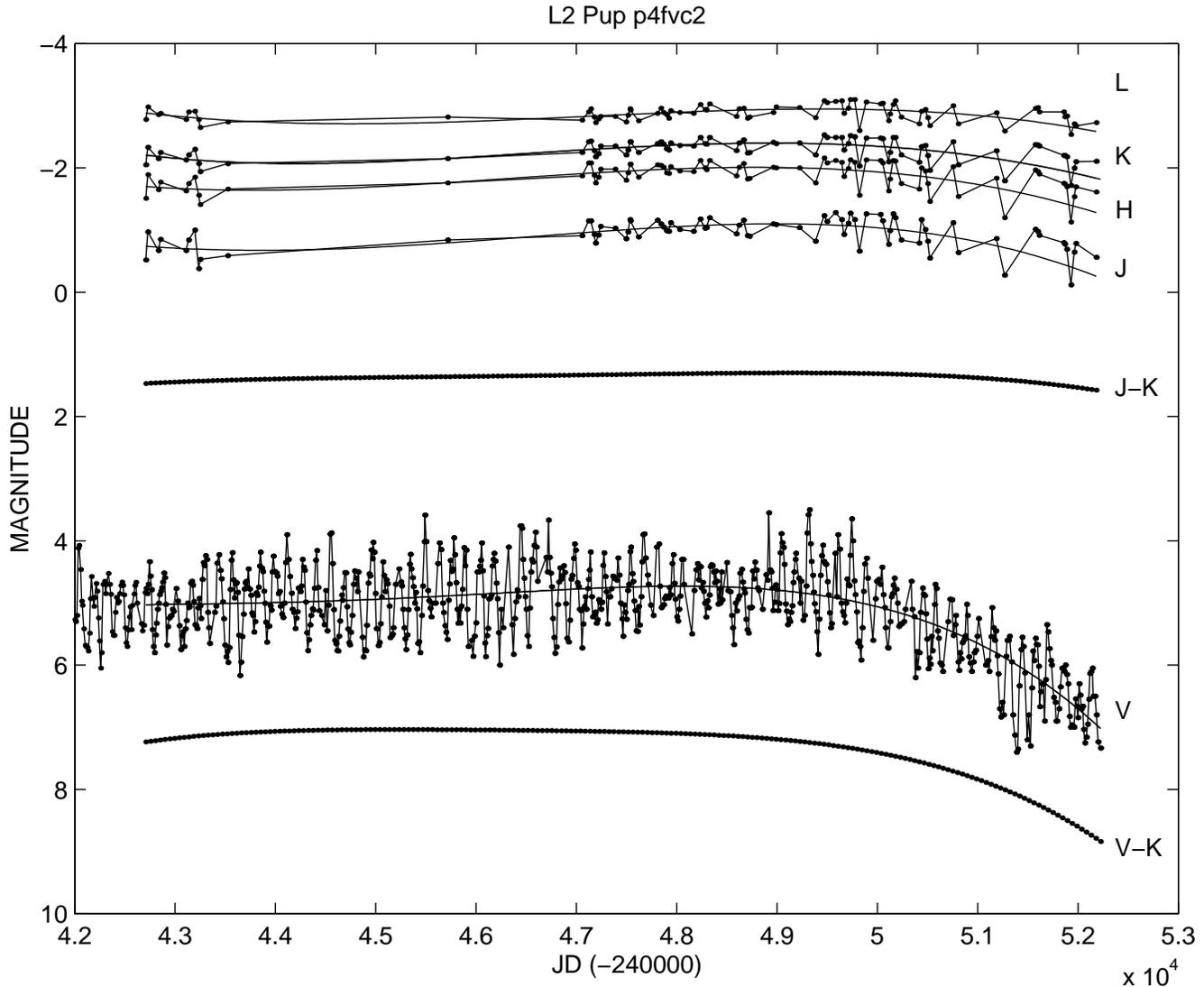}
\caption{\label{fig.ir} Infrared and visual photometry for \lpup.  }
\end{figure*}

We have taken $JHKL$ infrared photometry from \citet{WMF2000}, together
with more recent observations obtained with the same system (see
Table~\ref{table.ir}).  The infrared observations are much less frequent
than the visual ones and do not sample the pulsation cycle very well.
Nevertheless, the mean infrared magnitudes can be estimated, and we have
shown these in Fig.~\ref{fig.ir} by fitting a low-order polynomial to each
waveband.

\begin{table}
\caption[]{\label{table.ir} Near-infrared photometry of \lpup{} }
\begin{tabular}{ccccc}   \hline
JD $-$ 2400000 & $J$ & $H$ & $K$ & $L$ \\
\hline 
51186     & $-$0.865    & $-$1.833   & $-$2.276   & $-$2.889 \\
51239     & $-$0.273    & $-$1.200   & $-$1.792   & $-$2.594 \\
51574     & $-$1.013    & $-$1.968   & $-$2.375   & $-$2.958 \\
51599     & $-$0.976    & $-$1.941   & $-$2.369   & $-$2.972 \\
51613     & $-$0.918    & $-$1.902   & $-$2.352   & $-$2.901 \\
51858     & $-$0.796    & $-$1.753   & $-$2.201   & $-$2.901 \\
51868     & $-$0.775    & $-$1.742   & $-$2.199   & $-$2.823 \\
51887     & $-$0.691    & $-$1.695   & $-$2.178   & $-$2.834 \\
51928     & $-$0.118    & $-$1.130   & $-$1.720   & $-$2.538 \\
51962     & $-$0.645    & $-$1.536   & $-$1.998   & $-$2.708 \\
51979     & $-$0.783    & $-$1.696   & $-$2.101   & $-$2.676 \\
52181     & $-$0.562    & $-$1.614   & $-$2.106   & $-$2.730 \\
52242     & $-$0.601    & $-$1.497   & $-$1.964   & $-$2.632 \\
52258     & $-$0.877    & $-$1.764   & $-$2.138   & $-$2.776 \\
\hline
\end{tabular}
\end{table}

The dimming at $V$ is followed at all infrared bands, although with reduced
amplitude.  The $J-K$ colour has reddened during the dimming.

The mean $K$ magnitude before the dimming was $-2.34$ (Table~4 of
\citealt{WMF2000}).  More recently, the magnitude dropped to $-2.0$ or
fainter\footnote{\citet{B+Z98} gave $K=-2.65$, but this was a transcription
error and should have read $-2.15$.}.  The Hipparcos distance allows
comparison with the $K$-band period--luminosity (P--L) relation for nearby
Miras \citep{W+F2000}:
\begin{equation}
 M_K = -3.47 \log P +0.84.
\end{equation}
For \lpup{}, this relation predicts $M_K = -6.64$, which compares
relatively poorly to the observed value of $M_K = -6.16$.  The recent
fading has worsened the agreement.  Thus, \lpup{} appears to be one of a
small fraction of stars located below the P--L relation.  We will return to
this in Sec.~\ref{sec.pl}.

Table~\ref{table.photometry} shows the mean colours and magnitudes before
and after the dimming.  The pre-dimming NIR magnitudes are taken from
Table~4 of \citet{WMF2000}.  The current values are derived from the fits
in Fig.~\ref{fig.ir}, and are somewhat more uncertain because of the poor
sampling of the pulsation cycle.

\begin{table}
\caption[]{\label{table.photometry} Pre-dimming and present mean magnitudes
and colours of \lpup{} }
\begin{tabular}{ccc}   \hline
  & pre-dimming & present \\
 \hline
 $V$ &  ~4.8 & ~6.8 \\
 $J$ &  $-1.03$ & $-0.25$ \\
 $H$ &  $-1.94$ & $ -1.25 $ \\
 $K$ & $-2.34 $ & $ -1.8$ \\
 $L$ & $-2.91$ & $-2.7$ \\
 $V-K$ & $7.0 $ & $8.5$ \\
 $J-K$ & $1.31 $ & $1.55$ \\
 $m_{\rm bol}$ & ~0.73$^a$ & ~1.3$^a$ \\
\hline 
\noalign{\smallskip}
\noalign{$^a$ assuming no extinction correction}
\end{tabular}
\end{table}

\section{Discussion of the dimming}

What causes the large variations in mean magnitude of \lpup, especially the
current dramatic fading?  We consider three possibilities: changes in
temperature, luminosity and extinction.

\subsection{Change in stellar temperature and/or luminosity} 
\label{sec.temperature}

In late M-type giants, the optical flux is very sensitive to temperature.
Even a slight cooling would lead to a sharp drop in $V$ via the formation
of temperature-sensitive molecules (TiO, VO) that absorb strongly at
optical wavelengths \citep{R+G2002}.

We first need to estimate the effective temperature of \lpup.  In the
absence of a measured angular diameter, this requires reference to
published temperature scales.  The temperature scales for M-giants differ
greatly for stars with and without extended atmospheres.  Assuming no
extended atmosphere, the pre-dimming $V-K$ colour index of \lpup{} implies
$T_{\rm eff}\approx 3200\,\rm K$ \citep{BCP98,HBS2000}.  Models with this
temperature also give a $J-K$ colour consistent with the observed value.
\citet{JCP2002} adopted a temperature of $3400 \,\rm K$, also taken from
photospheric models, which is also consistent with the scale of
\citet{vBDB96}.

However, stars with extended atmospheres have much lower colour
temperatures \citep{Fea96,vBTCE2002}.  The following relation was given by
\citet{Fea96} for Miras, based on interferometric angular diameters:
\begin{equation}
 \log T_{\rm eff} = -0.474(J - K)_0 + 4.059.
\end{equation}
For \lpup, the pre-dimming $J-K$ colour index yields the much lower value
of $T_{\rm eff}=2775\,K$.
Since this star is Mira-like and shows evidence for mass loss, the
atmosphere is likely to be extended and the lower temperature is probably a
better description of its optical/NIR spectral energy distribution.

During the fading, $J-K$ increased to 1.55, which corresponds to a
temperature decline of 660\,K on the scale of \citet{Fea96}.  The increase
in $V-K$ to 8.5, would correspond to a temperature decline of about 300\,K
in the \citet{BCP98} models.  The bolometric correction, BC$_K$, increased
from 3.05 to 3.21 on the low-temperature scale, and from 3.05 to 3.15 on
the high temperature scale \citep{WMF2000,BCP98}.  The luminosity decline
in either case is about a factor of 1.8. Using $L \propto R^2 T_{\rm
eff}^4$, we conclude that the radius would have decreased by 10\%\ on the
high-T scale and increased by 20\%\ on the low-$T$ scale.

Such large changes in radius can be ruled out because of the stable period.
The maximum period change allowed by the observations is about~6\% (see
Fig.~\ref{fig.wavelet}).  Since the period of a pulsating star scales
inversely with the square root of stellar density, we can infer that the
radius cannot have changed by more than 4\% (since $P\propto R^{3/2}$).  In
fact, we can take the relative stability of the pulsation period over the
past 130 years to be evidence that the density of the star has not changed
significantly.  The period variations that are observed in \lpup{} are
similar to those seen in many semiregular variables and are most likely due
to the influence on the pulsation driving by random convective excitation
\citep{ChDKM2001}.  We discuss this in more detail in Paper~II
\citep{BKK2002}. 

The period is stable on the same timescales over which the mean magnitude
is variable, which argues that radius changes cannot be part of the
explanation of the dimming.  If we assume a constant radius, we could still
produce the observed change in $m_{\rm bol}$ if the temperature decreased
by about $400\,\rm K$.  Such a temperature change cannot be ruled out but
would require fine tuning.  Furthermore, such a large change in stellar
luminosity over a time scale of a decade only occurs in AGB stars in the
immediate aftermath of a thermal pulse.  Finding any star in such a
short-lived phase is very unlikely.  In any case, the thermal-pulse
scenario can also be ruled out because, during this phase, the radius
strongly contracts and the period shortens (e.g. \citealt{V+W93}), which is
not observed in \lpup{}.

It is therefore unlikely that either the radius or the luminosity of the
star has changed.  This points towards variable extinction as the cause of
the dimming.

\subsection{Extinction from circumstellar dust}

The plausible cause of the fading in \lpup{} is extinction by circumstellar
dust, which would not affect the amplitude or period of pulsation of the
underlying star.  Extinction laws for interstellar dust are well studied
and predict less extinction as one moves to longer wavelengths.  This is
certainly the case for \lpup, as can be seen from Fig.~\ref{fig.ir}: the
decline is much less dramatic in $JHKL$ than in~$V$.

Direct evidence for extinction comes from the present $J-K$ and $K-L$,
which are among the reddest of the M-type variables in the sample of
\citet{WMF2000}.  All stars with such red colours have $K$-band amplitudes
larger than 0.7\,mag, while the amplitude for \lpup{} is only 0.29\,mag at
$K$.  In fact, both $\Delta K$ and $\delta Hp$ (0.71) are small compared to
M-type variables.  This large discrepancy between colour and amplitude
suggests that the red colours are partly caused by extinction.

\lpup{} is also anomalously red for its period.  The average relation
between colour and period for Miras is
\begin{equation}
 J-K = -0.39 \pm 0.15 + (0.71 \pm 0.06) \log P,
 \label{eq.period-colour}
\end{equation}
\citep{WMF2000}.  For \lpup{}, this relation predicts $J-K = 1.13$,
compared with $1.31$ pre-dimming and $1.55$ at present.  The observed
values suggest that significant extinction existed even before the
present dimming. Assuming a linear relation between the $J-K$ excess
and the visual extinction, and assuming that the post-1995 dimming of
2\,mag at $V$ and 0.24 in $J-K$ is due to extinction, the pre-dimming
$J-K$ excess of 0.18 would correspond to a visual extinction of about
$1.5$\,mag at~$V$.

This estimate suggests that the long-term $V$-band variations may be
due to variable extinction.  The brightest epoch, in the 1950s, could
in this case give the best indication of the unreddened
magnitude.  However, while \lpup{} is Mira-like in some properties, it
is certainly not a true Mira and so results based on applying
Eq.~\ref{eq.period-colour} should be treated with some caution.

\begin{figure}
\includegraphics[bb=59 201 546 580,clip=true,
width=0.5\textwidth]{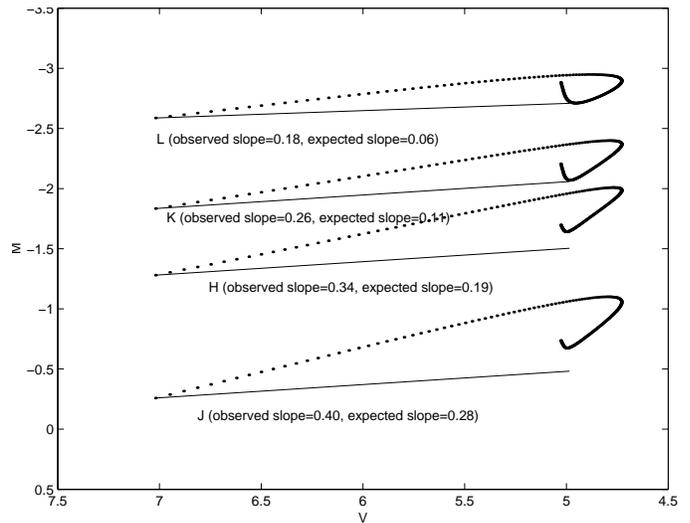}
\caption{\label{fig.extinction} Infrared versus $V$ photometry for \lpup,
using the polynomial fits shown in Fig.~\ref{fig.ir}.  The vertical axis is
the infrared magnitude, for $M = J, H, K, L$.  Written on each plot are the
observed slope and the expected slopes, the latter taken from Table~3 of
\citep{CCM89}. }
\end{figure}

\begin{figure}
\includegraphics[bb=59 201 546 580,clip=true,
width=0.5\textwidth]{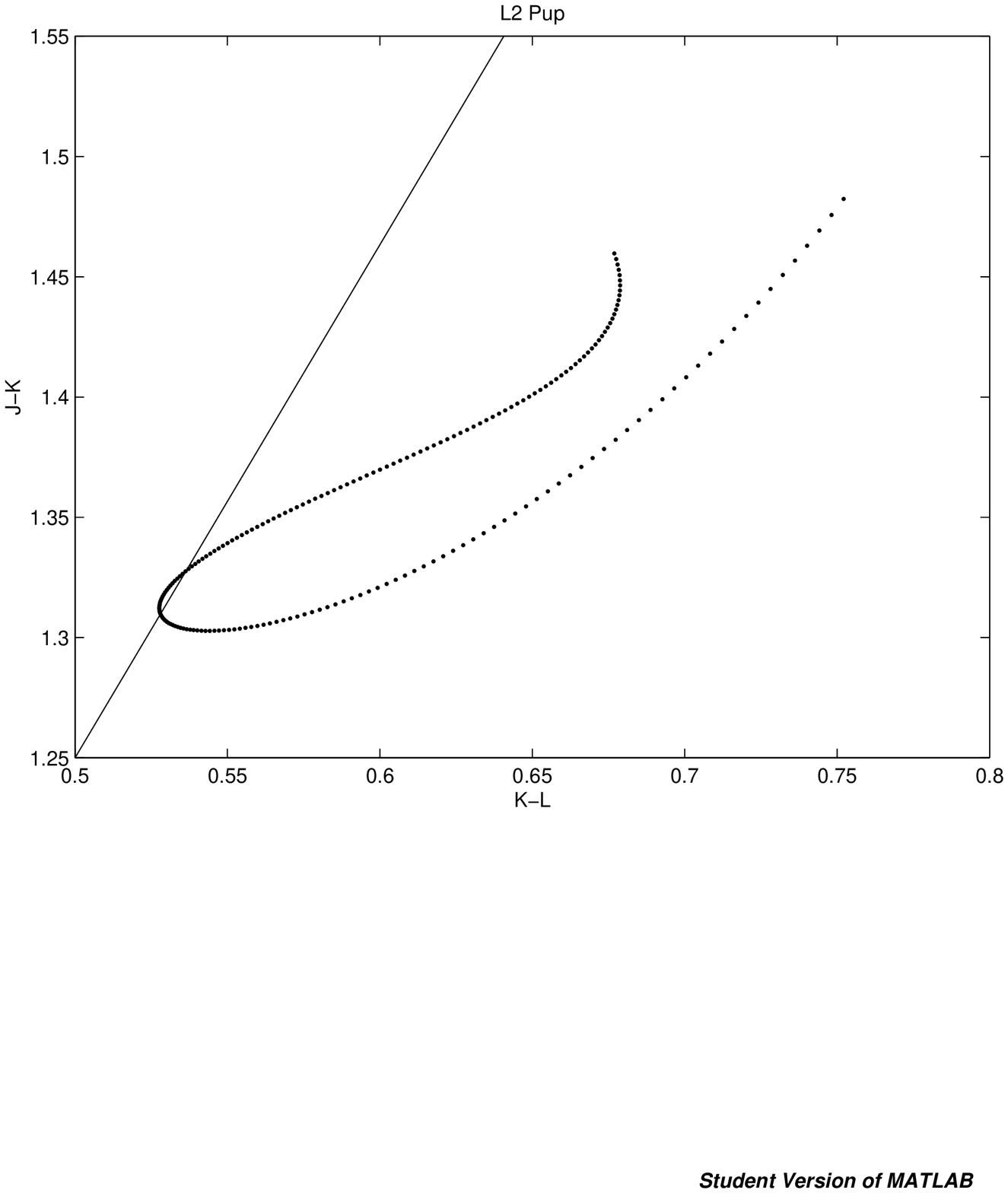}
\caption{\label{fig.colour} Infrared colour-colour diagram of \lpup, using
the polynomial fits shown in Fig.~\ref{fig.ir}.  The diagonal line shows
the interstellar reddening vector, with $A_V=1.0$, $E(J-K)=0.164$ and
$E(K-L)=0.045$.}
\end{figure}

\subsection{Extinction curve and the Mira P--L relation}
\label{sec.pl}

Typical extinction constants for interstellar dust are listed in
\citet{BCP98}.  For the observed post-1995 $J-K$ reddening, they give
$\Delta A_K=0.15$.  This is less than half the decline at $K$ shown in
Fig.~\ref{fig.ir}, suggesting this extinction curve is too steep.  To
investigate this quantitatively, we show in Fig.~\ref{fig.extinction} the
locus $(M,V)$ of the mean infrared magnitude versus the $V$ magnitude, for
$M = J, H, K, L$.  Also shown are straight lines with slopes taken from
Table~3 of \citet{CCM89}, which are standard values for extinction by
interstellar dust.  We can see that the observed slopes are in all cases
larger than expected, and the discrepancy is progressively greater at
longer wavelengths.  The interstellar extinction curve appears
inappropriate: instead the extinction appears to be `greyer.'  This could
be due to a contribution from oxides, which can form at high temperatures
\citep{Hen96}.  The infrared colour-colour diagram, shown in
Fig.~\ref{fig.colour}, confirms that the reddening is less steep than
expected from interstellar extinction.

The colour differential suggests that $A_{\rm K}/A_{\rm V} \approx 0.3$. If
we assume that the pre-dimming (1995) extinction at $V$ was $A_{\rm
V}\approx 1.5\,\rm mag$, the $K$-band would already have been affected by
0.5\,mag, so that $M_{\rm K,0} \approx -6.7$.  This is close to the value
expected from the Mira P--L relation, and well within the 1-$\sigma$
uncertainty contributed by the error on the Hipparcos parallax.  It
therefore appears possible that the Mira P--L relation is an accurate
indicator of the de-reddened magnitude of \lpup{}.

\subsection{Dust location}

For a pulsation velocity of 3\,km\,s$^{-1}$, the gas travels at most $1
R_\ast$ per year.  The short timescale of the dimming thus suggests that
the new dust is located within a few stellar radii of the star, possibly
less.  The rapid variation in polarization angle \citep{MCLG86} also
suggests that dust forms close to the star.  The timescale for dust
formation derived by these authors based on the polarization (about a
decade) is consistent with the timescale of extinction variations.

The polarization shows that the dust is not distributed isotropically, and
the variation in polarization angle shows that the distribution is not
constant.  Dust may form continuously, but in different directions: the
present dimming event would, in that case, correspond to dust formation
along the line of sight.

\subsection{Extinction episodes in other Mira variables}

Dimming episodes lasting a number of pulsation cycles (decades) are seen in
carbon Miras, with similarities to the dust obscuration events of R CrB
stars \citep{PCS90}.  \citet{LLoE97} found that these episodes are
accompanied by strong emission in the Swan C$_2$ band 5165\AA, while
\citet{WMF97} found them to occur in carbon stars with thick circumstellar
envelopes.

Could the fading of the non-carbon star \lpup{} be a related phenomenon?
\citet{LLoE97} listed five Miras with dust fadings which probably have
relatively low C/O ratios (W~Aql, U~Cyg, Y~Del, S~Lyr and RU~Vir), and
\citet{WMF2000} added Y~Vel to this list.  Unlike \lpup, all these stars
have long periods (433--490\,d).  And, in any case, inspection of the
visual light curves of all six stars on the AAVSO Web site do not show any
evidence for dimmings of the magnitude and longevity of those seen in
\lpup.  

Dimming episodes are seen in symbiotic Miras, where they are attributed to
variable dust extinction \citep{WHi87}.  A typical dimming at $J$ is 1--2
mag, even larger than that seen in \lpup.  We conclude that among
non-symbiotic oxygen-rich Miras and semiregulars, the dimming of \lpup{}
appears to be unique.

\section{Mass loss}

\citet{WMF2000} noted that \lpup{} stands out as one of only two
low-amplitude SRs with a large $K-[12]$ excess (the other being V~CVn),
implying a larger mass loss than is usual for SRs.  There may be an effect
from the low expansion velocity: the dust will remain close to the star
longer, thus increasing the [12]-$\mu$m excess.

Stronger evidence for on-going mass loss would come from a 10-$\mu$m
silicate feature.  Such a feature has not been reported, because of the
lack of a published 10-micron spectrum.  The lack of an original IRAS LRS
spectrum is common for bright stars (e.g., $o$~Ceti).  However,
\citet{V+C89} failed to recover the IRAS spectrum, which seems to have
suffered from a technical glitch.  ISO did not observe this star.

\begin{figure}
\includegraphics[angle=-90,width=0.45\textwidth]{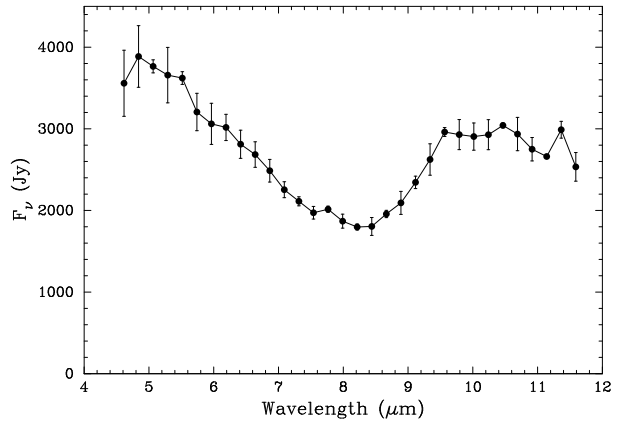}
\caption{\label{fig.MIRS} The mid-infrared spectrum taken by the MIRS on
board of the Japanese IRTS satellite. Open circles indicate channels
affected by glitches. The error bars give the 1-$\sigma$ scatter of
the signal in the off source region. Other sources of flux
uncertainty, for example, possible change of detector sensitivity, are
not included.}
\end{figure}

\lpup{} was included in the recent survey by the Japanese Infrared
Telescope in Space (IRTS) satellite \citep{MFG96}.  Observations were made
using the Mid-Infrared Spectrometer (MIRS) \citep{ROMcM94,YOT96}, which has
an $8\times8$ arcmin$^2$ aperture and scanned with a speed of
4\,arcmin/sec.  Spectra were taken on 1995 March~30 and April~5, i.e., just
before the dimming event.  We re-performed the data reduction to correct
for effects caused by the high brightness of the star.  The coverage in
this part of the sky by the IRTS was also below average, resulting in only
two well-centred scans over \lpup{}.  The averaged spectrum is shown in
Fig.~\ref{fig.MIRS}.

The spectrum shows a surprisingly strong silicate feature, with a
band-to-continuum ratio of about 2.5.  This give a silicate strength
parameter \citep{S+T89} of $A_{10} = 2.5 \log (F_{\rm obs}/F_{\rm cont})
\approx 1$.  The continuum between 6 and 8 $\mu$m has a slope consistent
with the stellar temperature but is  brighter than expected from
the $L$-band magnitude of $-2.9$.

There is no evidence for a large dust halo around \lpup{}. We inspected a 4
square degree field in the four IRAS images made using the IRAS Software
Telescope \citep{ABdJ95}: the field shows complicated cirrus structure but
no large envelope.

\subsection{Mass-loss rate}

The silicate feature is very strong compared to the limits on the
underlying dust continuum. To quantify this, we fitted the spectrum using
the dust model of \citet{S+K92}.  The model uses as input a blackbody,
which we assigned a temperature of $T_{\rm eff}=2900\,\rm K$ and luminosity
$L_\ast = 2.0 \times 10^3\,\rm L_\odot$. The dust was described as a
$r^{-2}$ distribution (i.e., constant mass-loss rate) with an inner and
outer radius: the mass-loss rate was converted to density assuming an
expansion velocity $v_{\rm exp}=2.5\,\rm km\,s^{-1}$.  The dust was
included as silicate grains with a size distribution $n \propto a^{-3.5}$,
with diameter $a>150$\,\AA.  In the fitting procedure, we adjusted the
stellar temperature and luminosity, the dust density (or mass-loss rate)
and inner radius, to reproduce the MIRS spectrum.  The stellar parameters
are largely determined by the shorter wavelength observations.

The 25 and 60-$\mu$m DIRBE measurements are much higher than the
corresponding IRAS values.  Part of this can be related to the different
response curves: both these DIRBE pass bands cut off a few microns bluer
than the IRAS bands.  However, the increase over the predicted IRAS flux
for the model spectrum is only about 10\%.  Although the DIRBE beam is much
larger than that of IRAS (as the telescope was only 19cm, compared to 60cm
for IRAS), but there are no other other bright point sources in the beam.

\begin{figure*}
\includegraphics[width=0.9\textwidth]{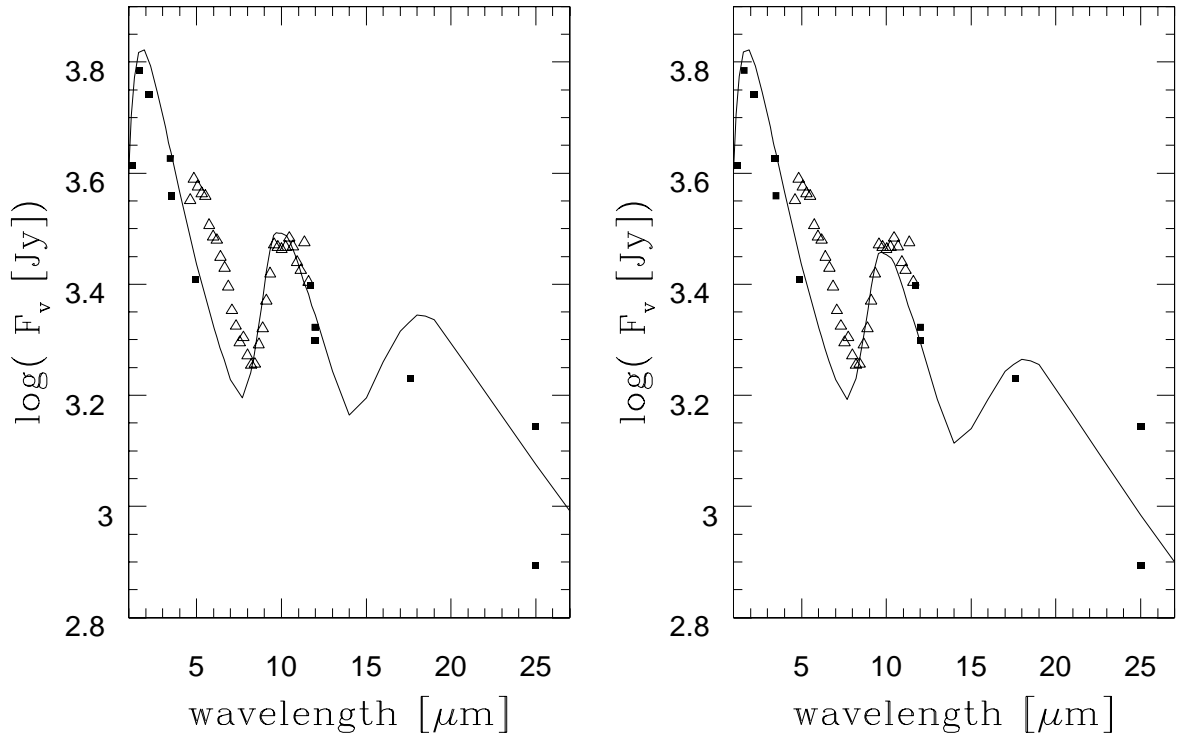}
\caption{\label{fig.ir.fit} Fit to the MIRS spectrum of \lpup{}.  The MIRS
spectrum (1995) is indicated by the open triangles. Filled squares are
the average $JHKL$ magnitudes, the IRAS 12 and 25$\mu$m fluxes (1985),
the DIRBE 3.5, 4.5, 12 and 25$\mu$m fluxes (1990), and the 11.7 and
17.6$\mu$m points (2001) of \citet{JCP2002}. Note that the 12 and
25$\mu$m points refer to very broad bands for which the effective
wavelength for this spectral energy distribution is shorter. The
left plot shows a fit for the silicate feature. The right plot 
attempts to better fit the longer-wavelength photometry}
\end{figure*}

It is difficult to reconcile the photometry with the silicate feature.
This is illustrated in Fig. \ref{fig.ir.fit}, where the left panel shows a
fit based the silicate feature, and the right panel shows a fit that
includes all photometry. The two models differ in their outer radius, which
is smaller for the right panel. The fitted parameters of the right panel
(our preferred model) are listed in Table~\ref{table.fit}.  The large
number of free parameters means that the fits are not unique: i.e., a
larger mass-loss rate with a thinner shell could also fit the spectrum.

\begin{table}
\caption[]{\label{table.fit} Fitted parameters to the infrared spectrum 
(1995: pre-dimming) of \lpup{}, corresponding to the right panel of
Fig.~\ref{fig.ir.fit}. }
\begin{flushleft}
\begin{tabular}{llllllllllll}   \hline
\hline
          &  Model       & IRAS     & DIRBE  & Keck \citep{JCP2002}\\
          &   (Jy)       &  (Jy)    &  (Jy)  &  (Jy) \\
\noalign{\smallskip}
 $F_{11.7}$ & 2120       &          &        & 2500 \\
 $F_{17.6}$ & 1810       &          &        & 1700 \\
 $F_{12}$  &  2100       &  1990    &  2137 \\
 $F_{25}$  &  1260       &   783    &  1385 \\
 $F_{60}$  &   127       &    88    &  151 \\
 $F_{100}$ &    22       &    27  \\
\hline
\end{tabular}
\begin{tabular}{ll}   
 Parameters:  & \\
$T_{\rm eff}$ & 2900\,K                   \\
$L_\ast$      & 2000\,$L_\odot$           \\
$r_{\rm in}$  & $7.2 \times 10^{14}$\, cm \\
$r_{\rm out}$ & $1.3 \times 10^{15}$\, cm \\
$T_{\rm d}^a$   & 350--200\,K              \\ 
$a_{\rm min}$  & 300\AA \\
$\dot M_{\rm d}$ & $5 \times 10^{-10}\,\rm M_\odot \,yr^{-1}$ \\
$\dot M_{\rm g}^b$ & $5.3 \times 10^{-7}\,\rm M_\odot \,yr^{-1}$ \\
\hline \\
\end{tabular}
\end{flushleft}
\vskip -10pt
{$^a$ range between smallest grains at inner radius and largest grains
 at outer radius}\\
{$^b$ assuming [Fe/H]$=-0.7$}
\end{table}

The most constraining aspects of the spectrum are the lack of dust emission
at 8$\mu$m and the strong silicate feature, which preclude the presence of
hot dust.  In our `best' model, the inner radius is at a dust temperature
of 350\,K, well below the dust condensation temperature.  The 17$\mu$m flux
(also dominated by silicate emission) is also surprisingly sensitive to the
inner radius and agrees with a dust shell detached from its condensation
point.  The outer radius is not as well determined as the inner radius.
However, a single thin shell can fit the long-wavelength IRAS/DIRBE flux
densities.  There is no evidence for a cold distant dust shell, consistent
with the low CO mass-loss rate found by \citet{OGDK2002}.

Between 4 and 8$\mu$m, the spectral energy distribution indicates a colour
temperature consistent with the photospheric temperature.  \citet{JCP2002}
argued that dust emission dominates from 5\,$\mu$m, based on evidence that
the variability behaviour is very different longward of 5\,$\mu$m.
However, the emission in this band in Mira variables is dominated by water
vapour at $\sim$2000\,K, located in the extended envelope at about 1.5
stellar radii \citep{MYC2002}.  This water vapour also raises the flux in
this region above that extrapolated from the photospheric $JHK$ magnitudes,
as seen in the fit.  Our findings do not indicate a substantial
contribution from dust to the spectrum shortward of 8\,$\mu$m.

We derive a (pre-dimming) mass-loss rate of $5 \times 10^{-7}\,\rm M_\odot
\,yr^{-1}$.  This is similar to that estimated by \citet{JCP2002}
(post-dimming).  They found a dust mass-loss rate 4 times larger than
derived by us because they adopted a dust outflow velocity of
10\,km\,s$^{-1}$ rather than 2.5\,km\,s$^{-1}$, but used a much lower
gas-to-dust ratio, which almost cancels this difference.  Our high
gas-to-dust ratio is based on the assumption that L$_2$ Pup is a thick-disk
star (see Sec.~\ref{sec.thickdisk}), for which we have assumed
[Fe/H]$=-0.7$.  In reality the gas mass-loss rate is not well determined.

\subsection{Silicate and Corundum}

The model shows that the photometric data imply a narrower silicate feature
than is observed.  The most likely explanation is that the observed
10-$\mu$m feature is not purely silicate.  In the scheme of
\citet{SBS2000}, this feature in \lpup{} would be classified either as
`broad' or possibly as `silicate D'. (The difference is subtle and requires
continuum subtraction, for which we lack sufficient wavelength coverage.)
They fitted the broad feature with a mixture of olivine (MgFeSiO$_4$) and
porous amorphous alumina (or corundum) (Al$_2$O$_3$).  Corundum has the
highest condensation temperature of any circumstellar solid, and is the
first condensate to form at $T\approx 1600$\,K.  Silicates may form
subsequently from gas-solid reactions with alumina \citep{Tie90}.
\citet{HAK97} suggested that the broad feature is seen in less evolved Mira
variables.  This may reflect the possiblity that, at low $\dot M$, the
conversion of alumina to silicate (replacing the Al with Mg) remains
incomplete, due to a low reaction rate.

We note that our model calculations used the pre-dimming spectrum.  It
would interesting to see whether the spectrum has changed significantly
since the dimming.  If alumina formed first, the peak of the 10$\mu$m
feature may have shifted to the red.  It is tempting to associate the
alumina with the grey extinction curve, but it has little extinction in the
optical (glass-like).  Better candidates would be TiO$_2$ or pyroxenes
\citep{Woi99}.  The precise agent responsible for the variable polarization
also remains undetermined.

\subsection{Evolutionary status}\label{sec.thickdisk}

\citet{JCP2002} pointed out that the luminosity and effective temperature
of \lpup{} do not unambiguously identify it as an AGB star, and allow the
possibility that it is on the first ascent red giant branch.  It is not
clear whether the spectrum of \lpup{} shows technetium, which is usually
taken as indicating an AGB star.  \citet{LLMB87} reported a probable
detection, while \citet{L+H99} listed only a possible detection.

However, if the $K$-band has significant obscuration then the bolometric
magnitude has been underestimated.  The bolometric correction would also
have been overestimated because of the observed $V-K$ is too large.  The
two effects combined would mean that the actual luminosity is substantially
greater than has been assumed, and is perhaps sufficiently large that
placement on the AGB is unambiguous.

Finally, as pointed out by \citet{JCP2002}, the space motion of \lpup{}
indicates it may be a thick disk star.  Given that the thick disk accounts
for up to 13\%\ of the local star density \citep{CSS2001}, it is not
unexpected to find a local Mira-like example.  If \lpup{} is indeed a
thick-disk star, as seems plausible but which cannot be stated with
certainty, the progenitor would likely have had a low metallicity and a low
mass.

\section{Conclusions}

Visual photometry of \lpup{} shows an unprecedented dimming over the past 5
years.  The long-term light curve shows stable periodicity, and we argue
that \lpup{} is Mira-like and should be classified as SRa.  The period
stability implies a constant stellar radius, which rules out temperature
and/or luminosity variations as the cause of the dimming.  Rather, the
dimming seems to arise from an episode of dust formation close to the
extended atmosphere.  Episodic dust obscuration events are fairly common in
carbon stars but have not been seen in (non-symbiotic) oxygen-rich stars.
We suggest that dust forms continuously but anisotropically, with the
current dimming event being due to to dust formation along the line of
sight.

The red colours indicate reddening from dust, but the extinction curve is
greyer than found for ISM dust.  This could reflect a higher fraction of
oxides.  The change of colour during the dimming indicates that already
before the dimming, the $V$-band magnitude was significantly affected by
circumstellar or atmospheric extinction.  \lpup{} was one of the few stars
located below the Mira $P$--$L$ relation: the derived $K$-band extinction
puts the star in closer agreement with this relation.  

We present a 10-$\mu$m spectrum showing strong silicate emission.  These
observations were carried out in 1995, just prior to the recent dimming.
The silicate feature can be fitted with a detached, thin dust shell, with
inner radius $7 \times 10^{14}\,\rm cm$ and outer radius $\approx 1.5
\times 10^{15}\,\rm cm$.  Longer wavelength photometry shows no evidence
for more distant, colder dust. We derive a mass-loss rate of $\dot M_{\rm
g} \approx 5 \times 10^{-7}\,\rm M_\odot \,yr^{-1}$, but this value depends
on the assumed expansion velocity and metallicity --- if the dust velocity
is high, the actual mass-loss rate could be higher.

\section*{Acknowledgments}

We are extremely grateful to the observers who have provided visual
observations, and to those who have maintained the databases and made them
publicly available.  We also thank Peter Tuthill for useful discussions.
TRB and AR are grateful to the Australian Research Council for financial
support.

\end{document}